\documentclass[prb,twocolumn,groupedaddress,epsfig]{revtex4}
\usepackage{graphicx}
\usepackage{setspace}

\newcommand{\bq}{\begin{equation}} 
\newcommand{\eq}{\end{equation}}
\newcommand{\bqa}{\begin{eqnarray}} 
\newcommand{\eqa}{\end{eqnarray}}
\newcommand{\nn}{\nonumber \\}
\newcommand{\ex}{\text{e}}

\begin{document}
\title{ A Fractionalized Quantum Spin Hall Effect}
\author{Michael W. Young}
\author{Sung-Sik Lee}
\author{Catherine Kallin}
\affiliation{Department of Physics and Astronomy,
         McMaster University,
         1280 Main St.~W, Hamilton, ON 
         L8S 4M1 
         Canada}
\begin{abstract}

Effects of electron correlations on a two dimensional quantum spin Hall (QSH) system are studied. 
We examine possible phases of a generalized Hubbard model on a bilayer honeycomb lattice with a 
spin-orbit coupling and short range electron-electron repulsions at half filling, based on the 
slave rotor mean-field theory. 
Besides the conventional QSH phase and a broken-symmetry insulating phase, we find a new phase,  
a fractionalized quantum spin Hall phase, where the QSH effect arises for fractionalized spinons which 
carry only spin but not charge.
Experimental manifestations of the exotic phase and effects of fluctuations beyond the saddle point 
approximation are also discussed.

\end{abstract}

\maketitle
\section{Introduction}
The quantum spin Hall (QSH) phase is a new state of matter which arises due to spin-orbit coupling in 
time-reversal symmetric systems\cite{KaneMele,Bernevig06}.
It is characterized by a gap in the bulk and an odd number of Kramers pairs of gapless edge modes 
which are protected by a $Z_2$ topological order\cite{KaneMele,Bernevig06,Roy06}.
Recently, an experimental signature for the gapless edge modes has been observed in HgTe quantum wells\cite{Bernevig}.
Although the QSH state was proposed in a non-interacting system\cite{KaneMele}, the gapless edge modes are 
stable in the presence of weak time-reversal symmetric disorder or many-body interactions\cite{CongjunWu06,Xu06}, 
which suggests that the topological order in the bulk is also robust 
 against weak disorder\cite{Essin07} and interactions\cite{Lee07,Qi,Ran}.
If electron correlations become sufficiently strong, a broken-symmetry insulating phase can be stabilized.
Recently, a possibility of the QSH effect arising from many-body interactions has also been studied\cite{Raghu,Grover}.

Fractionalized phases are another novel state of matter and arise as a result of electron correlations.
In the absence of spin-orbit coupling, a subtle balance between the kinetic energy of electrons and 
electron-electron interactions can stabilize spin liquid phases where spins remain disordered due 
to quantum fluctuations while charge excitations are gapped\cite{ANDERSON,PLEE}.  Spin liquids exhibit
fractionalization in that the low energy excitations are spinons which carry only spin but not charge.
Much attention has been paid to two-dimensional frustrated magnets which are candidates for spin liquid
states\cite{SHIMIZU1}.

Since either spin-orbit couplings or electron correlations can lead to an interesting phase, 
what happens when both of these interactions are important?
We address this question by examining the possibility of a new phase of matter arising due to an interplay between the spin-orbit coupling and electron correlations.
A \emph{fractional} QSH state which corresponds to a time-reversal symmetric version 
of the fractional quantum Hall state has been suggested as a possible phase 
for interacting systems with spin-orbit coupling\cite{Bernevig06}.
In this paper, we explore an alternative possibility 
where the QSH effect arises simultaneously with 
\emph{fractionalization} in a spin liquid state.
The honeycomb lattice is an ideal geometry to study such effects because it may support both the QSH  phase\cite{KaneMele} and the spin liquid phase\cite{Lee05,Hermele_su2}.

\section{The Model}\label{model}
We consider a generalized Hubbard model defined on a double layer of honeycomb lattice,
\begin{eqnarray}\label{hamiltonian}
 H & = -\sum_{( i,j ) \atop a,\sigma}
\left(t_{ija\sigma}c^{\dagger}_{ia\sigma}c_{ja\sigma} + \text{h.c.}\right)
+ U\sum_{i,a}\left(n_{ia}-1\right)^{2}  \nn & 
+ U^{'}\sum_{i}\left(n_{i1}-1\right)\left(n_{i2}-1\right)
- \sum_{ia}\mu_{a}\left(n_{ia}-1\right)
\end{eqnarray}
where 
 $c^{\dagger}_{ia\sigma}$ is the creation operator for an electron of spin $\sigma = \pm 1$ on 
site $i$ of layer $a=1$ or $2$, and $n_{ia}$ is the number operator.
$U (U^{'})$ is the on-site (interlayer) Coulomb repulsion and 
$\mu_{a}$ is the chemical potential which is tuned so that each layer is at half filling.
The intralayer tunneling amplitudes are $t_{ija\sigma} = t$ when $(i,j)$ are nearest neighbor (nn) sites
and $t_{ija\sigma} = \delta_{a1}t^{'}\text{e}^{i\phi_{ij}\sigma}$ for next nearest neighbor (nnn) sites.
We assume that there is no spin-orbit coupling or nnn hopping in the second layer and no interlayer tunneling.
The spin-dependent phase $\phi_{ij}\sigma$, which we 
take to be positive (negative) if an electron with spin up (down) hops around the lattice 
in a counter clockwise sense, is due to spin orbit coupling\cite{KaneMele}.
We emphasize that our model is an idealized model and the goal of our investigation
is to demonstrate the possibility of finding a new state of matter from the simplest 
model which contains both spin-orbit coupling and electron correlations.

We now represent the Hamiltonian in the slave rotor representation~\cite{Florens},
$c_{ia\sigma} = \text{e}^{-i\theta_{ia}}f_{ia\sigma}$, 
where the spinon operator, $f_{ia\sigma}$, carries only spin, 
and the chargon operator, $\text{e}^{i\theta_{ia}}$, carries only charge.
The enlarged Hilbert space is constrained by 
$L_{ia} = \sum_{\sigma}f^{\dagger}_{ia\sigma}f_{ia\sigma}-1$, 
where $L_{ia}= n_{ia}-1$ represents the charge quantum number, conjugate to $\theta_{ia}$.
Integrating out $L_{ia}$, appendix~\ref{bosonaction}, we obtain the partition function, 
$\mathcal{Z} = \int Df^{*}DfD\theta Dh\, \text{e}^{-\int d\tau L}$,
where the Euclidean Lagrangian is given by
\bqa
\label{action}
&& L  = \sum_{i,a\atop\sigma} f^{*}_{ia\sigma} \partial_{\tau} f_{ia\sigma} 
  +\sum_{i,a} (ih_{ia} - \mu_{a}) \left(\sum_{\sigma}f^{*}_{ia\sigma}f_{ia\sigma} -1\right) \nn
&& -\sum_{(i,j)\atop a,\sigma}
\left( t_{ija\sigma} f^{*}_{ia\sigma} f_{ja\sigma}
e^{i(\theta_{ia}-\theta_{ja})}+\text{h.c.}\right)  \nn
&& + \frac{1}{U_{+}} \sum_{i} ( \partial_{\tau} \theta_{i+} + h_{i+} )^2 
  + \frac{1}{U_{-}}\sum_{i}( \partial_{\tau} \theta_{i-} + h_{i-})^2.
\eqa
Here $U_{\pm}\equiv 2U\pm U^{'}$ and  
$A_{\pm}\equiv (A_{1}\pm A_{2})/2$ for $A_a = \theta_{ia}$ or $h_{ia}$,
where $h_{ia}$ is a Lagrange multiplier field enforcing the constraint.
In this paper, we concentrate on the parameter region  $U_- << U_+, t$, 
in which case the phase stiffness for the $\theta_{i-}$ field is large and 
the phases of the chargon fields in the two layers are locked together. 
If $\theta_{i-}$ is condensed, both $\theta_{i-}$ and $h_{i-}$ are gapped due to the Higgs mechanism.
At low energies, we can set the chargon fields and the Lagrange multipliers in the two layers 
equal to each other and our model reduces to the model with one chargon field, $\theta_{i}$, and one Lagrange multiplier, $h_i$.

We now decouple the quartic terms in the hopping sector by a Hubbard-Stratonovich transformation to obtain the effective Lagrangian,
\begin{widetext}
\bqa\label{hubbardaction}
&& L = \sum_{<i,j>} t\left[{\chi^{f}_{ij}}^{*}\chi^{X}_{ij} + \text{h.c}\right]
 + \sum_{<<i,j>>} t^{'}\left[{{\chi^{f}_{ij}}^{'}}^{*}{\chi^{X}_{ij}}^{'} + \text{h.c}\right] 
  + \sum_{i,a}\mu_{a}+\sum_{i}\left(\bar{\lambda}_{i} + \sum_{a} \bar{h}_{ia}\right)
  \nn
&&  + \sum_{i,a,\sigma}f_{ia\sigma}^{*}(\partial_{\tau} -\bar{h}_{ia}-\mu_{a})f_{ia\sigma} 
   -\sum_{<i,j>\atop a,\sigma}t\left[\chi^{X}_{ij}f^{*}_{ia\sigma}f_{ja\sigma} + \text{h.c}\right] 
  -\sum_{<<i,j>>\atop \sigma}
t^{'}\left[ {\chi^{X}_{ij}}^{'} \text{e}^{i\phi_{ij}\sigma}f^{*}_{i1\sigma}f_{j1\sigma} + \text{h.c}\right]  \nn
&&
 +  \frac{1}{U_+}\sum_{i}\left(\partial_{\tau} + \bar{h}_{i}\right)X_{i}^{*}
       \left(\partial_{\tau} -\bar{h}_{i}\right)X_{i}  
   - \sum_{<i,j>}t\left[\chi^{f}_{ij}X^{*}_{i}X_{j} + \text{h.c}\right]  
 -  \sum_{<<i,j>>}t^{'}\left[{\chi^{f}_{ij}}^{'}X^{*}_{i}X_{j} + \text{h.c}\right] \nn
&&
- \sum_{i} \bar \lambda_{i}|X_{i}|^{2}.
\label{4}
\eqa
\end{widetext}
Here a soft boson field $X_{i}\equiv\text{e}^{-i\theta_{i}}$ has been introduced with a Lagrange multiplier $\lambda_{i}$ which imposes the constraint $|X_{i}|=1$.
$\bar{\lambda}_{i} = -i \lambda_{i}$ and $\bar{h}_{i}= -i h_i$ are the saddle point values of the Lagrange 
multipliers and lie on the imaginary axis\cite{Lee05}.
$\chi^{f}_{ij}$ and $\chi^{X}_{ij}$ ($\chi^{f'}_{ij}$ and $\chi^{X'}_{ij}$) are the nn (nnn) hopping order parameters of spinon and chargon respectively.

\section{Mean-field Phase Diagram}
In the small $U$ limit, the system essentially reduces to a non-interacting model with no coupling between the layers. 
In this limit, the conventional QSH phase will be realized in the first layer where there is a spin-orbit coupling.
In the second layer, the semi-metal (SM) phase with gapless Dirac fermions will be obtained.

When $U>>~t,t^{'}$, the Coulomb interactions are dominant and the low energy states of the system are 
described by the configurations which satisfy $\sum_{a}n_{ia}=2$.
To second order in $t$ and $t^{'}$, the low energy effective Hamiltonian is obtained to be
\bqa
\label{effectiveham}
H_{\rm{eff}} &=& \frac{t^{2}}{U}\sum_{<i,j>}\text{tr}\left[\mathcal{Q}_{i}\mathcal{Q}_{j}\right] \nn
& + & \frac{{t^{'2}}}{U}\sum_{<<i,j>>}\text{tr}
\left[\text{e}^{-i\phi_{ij}\sigma^{3}}\mathcal{T}_{i}\text{e}^{i\phi_{ij}\sigma^{3}}
\mathcal{T}_{j}\right],
\label{es}
\eqa
where $[\mathcal{Q}_{i}]_{a\sigma ,b\sigma^{'}}=c^{\dagger}_{ia\sigma}c_{ib\sigma^{'}}$ 
is the $4 \times 4$ matrix of U(4) generators
and 
$[\mathcal{T}_{i}]_{\sigma,\sigma^{'}}=c^{\dagger}_{i1\sigma}c_{i1\sigma^{'}}$  is the $2 \times 2$ 
matrix of U(2) generators which are restricted to the first layer.
The first term has a $U(4)=U(1)\otimes SU(4)$ symmetry where the $U(1)$ is associated with conservation of the 
total charge and the $SU(4)$ with conservation of the flavor quantum number given by the layer index and the spin.
The $6$ states which satisfy the constraint $\sum_{a}n_{ia}=2$ at each site form the rank $2$ anti-symmetric  
representation of the $SU(4)$ group.
The second term breaks the $U(4)$ symmetry into $SU(2)\otimes U(1)^3$ where the unbroken $SU(2)$ symmetry 
is the spin-rotational symmetry in the second layer and the three U(1) symmetries are associated with 
charge conservation in each layer and $S_z$ conservation in the first layer.
If $t^{'} = 0$, each nn bond tends to form an $SU(4)$ singlet 
and a valence bond solid (VBS) phase which breaks translational symmetry is a good candidate for the ground state\cite{Affleck88}.
The most natural pattern among possible VBS states in the honeycomb lattice is the dimerized phase where 
valence bonds are stronger for the bonds which are directed along one of the six symmetry directions\cite{Lee05}.
A non-zero $t^{'}$ will enhance quantum fluctuations, but we expect that the fully gapped dimerized state will remain stable for a finite range of $t^{'} < t$.

With the guidance of these insights, the mean field theory 
is carried out for the uniform and dimer ansatze.
We solve a system of self-consistent equations at $T=0$ for 
the link order parameters, chemical potentials and Lagrange multipliers, 
by requiring that the energy remains stationary with respect to variations of those variables.
We then find the mean-field  phase diagram by choosing the 
lower energy configuration between the dimer and uniform ansatze. 
In particular, we are interested in finding a new phase in the insulating side 
of the phase diagram where both the spin-orbit coupling and the electron correlation are important.
Although, we could start from an effective `spin' model to study such insulating phases,
we will use the full action in Eq. (\ref{4}) which is applicable in all parameter regimes.
It would be of interest to study the possibility of obtaining an exotic phase 
in an effective model like Eq. (\ref{es}), possibly with additions of higher order ring-exchange terms.

\begin{figure}
\includegraphics[scale=0.5,width=3in,trim=0in 0.1in 0in 0.0in]{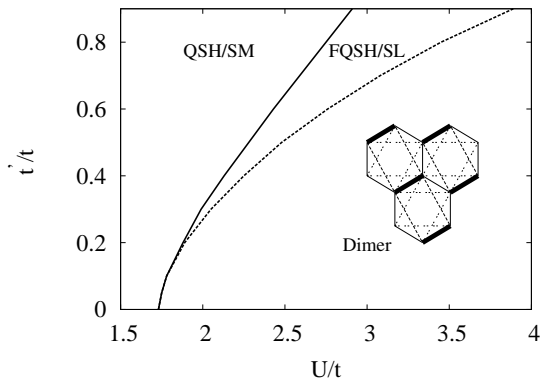}
\caption{\label{phasediagram} 
Phase diagram in the space of $t^{'}/t$ and $U/t$  in a $40 \times 40$ lattice with $U_-=0$.  
The weakly interacting phase (small $U$) has $Z \neq 0$ and the first layer forms the conventional QSH phase 
while the second layer is in the semi-metal (SM) phase with gapless Dirac nodes. 
The intermediate region has the fractionalized quantum spin Hall (FQSH) phase with $Z=0$
where chargeless spinons form the QSH phase in the first layer and 
the gapless spin liquid (SL) phase in the second layer.
In both QSH/SM and FQSH/SL phases, the nn and nnn hopping order parameters are nonzero and site-independent.
The large $U$ region is a dimerized phase where the hopping order parameters along the bold lines 
have the maximum amplitude and all other bonds have zero amplitude.  
The solid line represents the second order transition
and the dotted line, the first order transition.
}
\end{figure}

For large $U$, we find the dimerized configuration has lower energy, while for small $U$ the 
uniform configuration has lower energy, as expected.
There is a first order phase transition between these two phases.
Within the uniform phase, an onset of the chargon condensation marks another phase transition.
Although not shown here, the chargon gap vanishes continuously 
as $U$ decreases and the phase transition is a second order phase transition.
The Bose condensation amplitude is given by $Z = |< X >|^2$. 
If the chargeons are condensed, a spinon recombines with a chargon to become an electron.
This phase is the conventional weakly interacting phase where the electrons form the QSH phase in the first layer 
while the semi-metal (SM) phase with Dirac nodes is realized in the second layer.
For $t^{'}/t < 0.2$, the first order phase transition from the uniform phase to the dimerized phase 
occurs before the Bose condensation amplitude $Z$ becomes zero as $U/t$ increases so that there is no intermediate phase between the conventional QSH/SM phase and the dimerized phase.
On the other hand, for $t^{'}/t > 0.2$, a window opens up for an intermediate phase and the region of stability for
the intermediate phase grows as $t^{'}$ is increased.
The mean-field phase diagram is shown in Fig. \ref{phasediagram}.\newline
\indent The intermediate phase is characterized by the uniform link order parameters but, unlike the QSH/SM  phase, 
the chargeons remain gapped, which makes it an insulating phase. 
This is a phase where the fractionalized spinon arises as a low energy excitation.
In the first layer, the spinon is gapped in the bulk due to the the spin dependent phase in the spinon 
hopping which has been inherited from the spin-orbit coupling of electrons as is shown 
in the seventh term of Eq. (\ref{4}).
At the mean-field level, which ignores the fluctuations of the order parameters, the spinon spectrum 
is essentially the same as the electron spectrum in the Kane and Mele model~\cite{KaneMele}.
The non-trivial topological structure in the spectrum guarantees that there exist gapless edge modes in the first layer.
Therefore, we have a \emph{fractionalized quantum spin Hall} (FQSH)  phase in which the gapless edge states are 
carried by spinons and not by electrons as in the conventional QSH phase.
It is noted that the gapless edge mode and the FQSH state may be robust even though $S_z$ symmetry is broken in the first layer as will be discussed in the next section.
In the second layer, the spins form an algebraic spin liquid (SL)\cite{WEN}, whose low energy excitations are described by  four two-component Dirac spinons.\newline
\indent The electromagnetic response and transport properties of the FQSH phase are very different from those of the usual QSH phase, as discussed below in Sec. V.
It was recently pointed out that the conventional QSH state can have spin-charge separated excitations 
in the presence of $\pi$-flux even in the absence of many-body interactions\cite{Qi}.
We emphasize that the spinon which arises in the FQSH phase is different in that 
they are intrinsic excitations resulting from many-body correlations while the fractionalized 
excitations obtained in the non-interacting systems are generated by an external fractional magnetic flux quantum. 
The QSH effect in the presence of a $Z_2$ gauge field was recently studied 
where the dynamic fluxon makes the fractionalized excitation a propagating mode\cite{Ran}.

\section{Stability of the edge modes}
Beyond the mean-field approximation, the most important fluctuations are the phase fluctuations 
of the hopping order parameters.
The phase mode is described by a gauge field because it restores the gauge invariance associated with the local phase transformation $f_{ia\sigma} \rightarrow e^{i\varphi_i} f_{ia\sigma}$ and $\theta_i \rightarrow \theta_i + \varphi_i$.
The low energy effective theory in the FQSH/SL phase is given by
\bqa\label{lowenergy}
S & = &  \sum_{n,\sigma}  \int d\tau dx_1 dx_{2}~ \bar{\psi}_{n \sigma}(i\gamma^{\mu}D_{\mu})\psi_{n \sigma} \nn
& + & \frac{1}{g^2} \int d \tau dx_1 dx_2 ~ f_{\mu \nu} f_{\mu \nu} 
+ \int d\tau dx_1  ~ \bar{\eta}( i\gamma^{a}D_{a} ) \eta. 
\label{ea}
\eqa
Here  $\psi_{n \sigma}$ is the 2+1D massless Dirac fermion in the second layer,
$\sigma$ labels the Kramers doublet which corresponds to spin in the $S_z$ conserved case   
and $n=1,2$ is the index for the nodal points.
$D_{\mu}= \partial_{\mu}-ia_{\mu}$ is the covariant derivative, $a_\mu$  is the internal gauge field and $f_{\mu\nu}$ is the field strength tensor with $\mu=0,1,2$.
$\eta$ is the 1+1D Dirac fermion on the edge of the first layer with $a=0,1$.
The edge is assumed to be along the $x_1$ direction.
Although the gauge field is a compact U(1) gauge field, the compactness is unimportant at low energies when $S_z$ is conserved\cite{RanSenthil} or  a large number of gapless Dirac fermions are coupled with the gauge field\cite{Hermele}.
In our case, there are $N=4$ gapless Dirac fermions coming from the second layer.
In the following, we proceed with the assumption that the four gapless Dirac fermions are 
enough to stabilize the fractionalized phase against proliferation of instantons. 
It is noted that the stability of the FQSH state relies on the existence of both layers.
The spin dependent nnn hopping in the first layer opens up the gap of the spinon in the first layer which provides the robustness of the edge modes. 
The presence of the second layer is crucial in that the gapless spinons screen the gauge field and suppress the gauge fluctuations.\newline
\indent The U(1) gauge field is coupled to the spinons in both layers.
The spinons are gapped in the bulk of the first layer but there are gapless edge modes.
Although the existence of the gapless modes has been inferred from the mean-field band structure which 
has a non-trivial topological order, the stability of those edge modes is less clear in this case because 
they are coupled to the gapless gauge field.
The key question is whether the fluctuating gauge field destabilizes the topological order associated with the spinon band to open up a gap for the edge modes.
In order to address this question, one can integrate out the bulk degrees of freedom in Eq.~(\ref{ea}) to obtain an effective action at the edge.
The resulting theory is an 1+1D quantum electrodynamics (QED) with a non-local action for the gauge field.
Whatever this non-local action is, in the 1+1D QED the quantum fluctuations of the 
gapless fermions open up a gap for the gauge field\cite{Schwinger}.
This suppresses the fluctuations of the gauge field at the edge although the gauge field remains gapless in the bulk.\newline

\indent 
One may worry about the possibility of direct spin-spin interactions between the two layers destabilizing the edge modes.
To examine the stability of the edge modes, one has to consider all the gapless modes in the low energy theory  (\ref{lowenergy}).
Since there is no tunneling between the two layers, the lowest order inter-layer interactions that one can add are two-body terms of the form
\begin{equation}\label{interactions}
V\int d\tau dx_1~\bar{\psi}(\tau,x_1,x_2=0)\psi(\tau,x_1,x_2=0)\bar{\eta}(\tau,x_1)\eta(\tau,x_1)~.
\end{equation}
Since the edge modes in the first layer can only interact locally with the bulk modes in the second layer the integration measure only has one spatial and one temporal component.  
Neglecting gauge fluctuations and forward scatterings of the edge modes, 
the free low energy theory is invariant under a scale transformation 
$(\tau,x_1,x_2)= b (\tau^{'},x_1^{'},x_2^{'})$, $\psi= b^{-1}\psi^{'}$ and 
$\eta= b^{-1/2}\eta^{'}$ with $b>1$.
The inter-layer interaction scales as $V^{'}=b^{-1}V$.
If we include gauge fluctuations and forward scatterings of the edge modes,
the edge mode is described by the Luttinger liquid with a nontrivial Luttinger parameter $K \neq 1$
and the spinons in the second layer are described by the algebraic spin liquid.
As a result, the scaling dimension of the inter-layer coupling will receive loop corrections
which are of the order of $[V]= -1 + O(1/N) + O(K-1)$.
Given that $N=4$, the inter-layer coupling may remain irrelevant if the forward scattering is sufficiently weak.

\begin{figure}
\includegraphics[width=3in,trim=0in 5.5in 0in 1in]{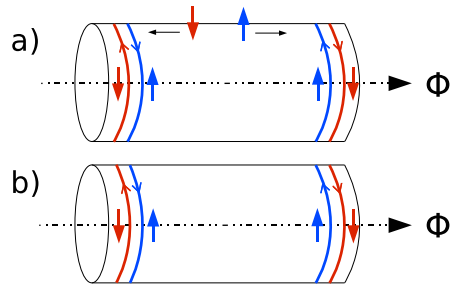}
\caption{\label{fluxtube} 
(a) Transverse spin response to an applied external magnetic field in the conventional quantum spin Hall phase. 
Upon threading a magnetic flux quantum, a spin up propagates from one edge 1, say, to edge 2 and a spin down propagates from edge 2 to edge 1.  
(b) The response in the fractionalized quantum spin Hall phase.
The external flux does not generate any transverse spin transport because the edge modes are charge neutral spinons.
}
\end{figure}

\section{Physical Properties and discussion}
Now we discuss physical manifestations of the FQSH state.
The longitudinal transport properties along the edge are very different from those of QSH states or trivial insulators.
There will be a metallic thermal conductivity along the edge due to the gapless edge mode.
However, there will be no charge conductivity because the spinon is charge neutral, which is the signature of the spin-charge separation.\newline
\indent The most stark difference from the conventional QSH state lies in the transverse spin transport induced by an external electromagnetic (EM) field.
We put the system on a cylinder with two edges at the ends of the cylinder.
In the usual QSH state with $S_z$ conservation, 
upon threading a magnetic flux quantum through the halo of the cylinder, 
a spin up electron is transported from one edge to the other 
while a spin down electron is transported in the opposite direction.
This results in a transport of net spin $S=1$ from one edge to the other.
This is illustrated in Fig.~\ref{fluxtube}(a).
On the other hand, in the FQSH phase, the edge modes are neutral spinons which are not directly 
coupled to the external EM field and there will be no such transverse spin transport.
Although spinons are indirectly coupled to external EM fields through chargeons, which are coupled to both the 
external and internal gauge fields, the weak coupling cannot produce a nonzero spin Hall transport because of a non-trivial quantum order associated with the fractionalization.
In the fractionalized phase, the tunneling rate of the internal gauge flux from one value to another value is exponentially suppressed with increasing system size and the flux through the cylinder is precisely conserved at $T=0$ in the thermodynamic limit.
The internal gauge flux remains strictly at zero under the adiabatic insertion of the flux.
Therefore, the external flux does not induce any transverse spin transport, as is illustrated 
in  Fig.~\ref{fluxtube}(b), in sharp contrast to the QSH state.
This insensitivity of the edge modes to EM fields can potentially be useful in stabilizing the edge modes in an environment with fluctuating EM fields which induce back scatterings between the edge modes in QSH states.

In summary, we proposed and studied a simple model which has both spin-orbit coupling and many-body interactions.
We found a region of the mean-field phase diagram where a fractionalized quantum spin hall (FQSH) state is stable 
and argued that this state may survive the effects of fluctuations under certain conditions.
In the FQSH state, charge neutral spinons form gapless edge modes which carry only spin, unlike the conventional QSH state where the edge modes carry both charge and spin.
Due to the charge neutral edge modes, the FQSH state shows a set of unique transport properties and electromagnetic responses which are distinct from conventional states.

\section{Acknowledgments}
We acknowledge useful discussions with Y. Ran and T. Senthil.
This work was supported by NSERC (C. K. and S. L.) and by CIFAR (C. K.).

\appendix
\section{Derivation of the Boson Action in Eq. (\ref{action}) }\label{bosonaction}

Since the constraint $L_{ia}= n_{ia}-1$ is diagonal in site indices calculating its partition function can be reduced to calculating one site 
matrix elements of the form
\begin{equation}
       \mathcal{Z}_{i\theta} = \langle\theta^{'}_{1}\theta^{'}_{2}|\ex^{-\epsilon H_{L}}|\theta_{1}\theta_{2}\rangle
\end{equation}
where $1$ and $2$ refer to the layer index.  Here $H_{L} = U(L^{2}_{1} + L^{2}_{2}) + U^{'}L_{1}L_{2} - i(h_{1}L_{1} + h_{2}L_{2})$.  The one site 
partition function becomes
\begin{equation}\label{onesiteangularpartitionfunction}
       \mathcal{Z}_{i\theta} = \sum_{l_{1},l_{2}}\ex^{il_{1}(\theta^{'}_{1}-\theta_{1})+il_{2}(\theta^{'}_{2}-\theta_{2}) -\epsilon\left[
                                       U(l^{2}_{1} + l^{2}_{2}) + U^{'}l_{1}l_{2} - i(h_{1}l_{1} + h_{2}l_{2})\right]}~,
\end{equation}
where we have omitted the site dependence of the eigenvalues to simplify the forthcoming formulae.\newline
\indent To obtain the effective action of the $\theta_{ia}$ variables we must sum over all $l_{ia}$. 
 We do this by making a change of variables from the discrete $l_{ia}$ to a new set of `continuous' variables $p_{a} = \epsilon l_{a}$.  After implementing 
these changes we then change variables from the original $p_{1}$ and $p_{2}$ to new symmetric and antisymmetric variables
\begin{equation}
     p_{\pm} \equiv \frac{1}{2}\left(p_{1} \pm p_{2}\right),
\end{equation}
which allow us to write the one site partition function as two decoupled Gaussian integrals
\begin{eqnarray}
 \nonumber      \mathcal{Z}_{i\theta} &=& \frac{1}{2\epsilon^{2}}\int dp_{+}dp_{-}\\
       \nonumber          &&\times\ex^{2ip_{+}\left[\dot{\theta}_{+}+h_{+}\right]  +  2ip_{-}\left[\dot{\theta}_{-}+h_{-}\right]
                      -\frac{1}{\epsilon}\left[(2U+U^{'})p^{2}_{+} + (2U-U^{'})p^{2}_{-}\right]},\\
\end{eqnarray}
where we have rewritten all fields as symmetric and antisymmetric combinations 
of the original layer dependent fields.  
Defining new coupling constants 
as $U_{\pm} = 2U \pm U^{'}$,
we obtain
\begin{equation}
      \mathcal{Z}_{i\theta} = \frac{1}{2\epsilon^{2}}\sqrt{\frac{\pi\epsilon}{U_{+}}}\sqrt{\frac{\pi\epsilon}{U_{-}}}~
       \ex^{-\frac{\epsilon}{U_{+}}(\dot{\theta}_{+} + h_{+})^{2}}
                     \ex^{-\frac{\epsilon}{U_{-}}(\dot{\theta}_{-} + h_{-})^{2}}~.
\end{equation}
\indent 
The full partition function for the $\theta$ variables is obtained by taking 
a product over all lattice sites of the single site result above.   
This gives the last two terms in Eq.~(\ref{action})

\section{Exactness of one-Boson theory when $U^{'}=2U$}
In section~\ref{model} we argued that in the region $U^{'}\approx 2U$ our model 
reduces to the one-boson model through the Higgs mechanism.  
Here we show that the one-boson model becomes exact when $U^{'}=2U$.
For $U^{'}=2U$, we can write Hamiltonian~(\ref{hamiltonian}) as
\begin{eqnarray}\label{specialhamiltonian}
\nonumber H  &=& -\sum_{( i,j ) \atop a,\sigma}\left(t_{ija\sigma}c^{\dagger}_{ia\sigma}c_{ja\sigma} + \text{h.c.}\right)\\
\nonumber &&+ U\left(\sum_{\alpha=1}^4 c^{\dagger}_{i\alpha}c^{\phantom{\dagger}}_{i\alpha}-2\right)^{2}\\
 &&- \sum_{ia}\mu_{a}\left(n_{ia}-1\right),
\end{eqnarray}
where we have introduce an $SU(4)$ index, $\alpha=1,..,4$, defined as $1=(1\uparrow)$, $2=(1\downarrow)$, $3=(2\uparrow)$, and $4=(2\downarrow)$; 
the first letter in the parenthesis is the layer index and the arrows represent the eigenvalue of $S_{z}$.  The Coulomb term is now 
an $SU(4)$ symmetric interaction term.\newline
\indent We can now decompose the electron operator into a spinon part and chargeon part as $c_{i\alpha}=f_{i\alpha}\ex^{-i\theta_{i}}$, where 
the $SU(4)$ quantum number is carried by the spinon. 
With this decomposition we obtain the slave-rotor representation for an $SU(4)$ model\cite{Florensoring},
\begin{eqnarray}\label{SLspecialhamiltonian}
\nonumber H  &=& -\sum_{( i,j ) \atop a,\sigma}
 \left(t_{ij\alpha}f^{\dagger}_{i\alpha}f^{\phantom{\dagger}}_{j\alpha}\ex^{i (\theta_{i}-\theta_{j})} + \text{h.c.}\right)
+ U\sum_{i}L_{i}^{2}\\ 
\nonumber &&+i\sum_{i}h_{i}\left(\sum_{\alpha}f^{\dagger}_{i\alpha}f^{\phantom{\dagger}}_{i\alpha}-L_{i}-2\right)\\
 && - \sum_{i,\alpha}\tilde{\mu}_{\alpha}\left(f^{\dagger}_{i\alpha}f^{\phantom{\dagger}}_{i\alpha}-\frac{1}{2}\right).
\end{eqnarray}
Here $h_i$ is the Lagrange multiplier which imposes the constraint 
$L_{i}=\sum_{\alpha}f^{\dagger}_{i\alpha}f^{\phantom{\dagger}}_{i\alpha}- 2$ 
with $L_i$ being the conjugate variable to $\theta_i$.
We have defined a new chemical potential $\tilde{\mu}_{\alpha}=\mu_{1}$ if $\alpha=1,2$ and $\tilde{\mu}_{\alpha}=\mu_{2}$ if $\alpha=3,4$.
If we apply the similar Hubbard-Stratonovich transformation to this Hamiltonian we would reproduce the effective action in Eq.~(\ref{hubbardaction}).

\end{document}